# High-Dimensional Operator Learning for Molecular Density Functional Theory

Jinni Yang[1], Runtong Pan[2], Jikai Sun[2], and Jianzhong Wu[2*]

[1]*School of Physics, Jilin University, Changchun, Jilin 130015, P. R. China*

[2]*Department of Chemical and Environmental Engineering, University of California,*

*Riverside, CA 92521, USA*

**Abstract**

Classical density functional theory (cDFT) provides a systematic approach to predict the structure and thermodynamic properties of chemical systems through the single-molecule density profiles. Whereas the statistical-mechanical framework is theoretically rigorous, its practical applications are often constrained by challenges in formulating a reliable free-energy functional and the complexity of solving multidimensional integro-differential equations. In this work, we established an optimized operator learning method that effectively separates the high-dimensional molecular density profile into two lower-dimensional components, thereby exponentially reducing the vast input space. The convoluted operator learning network demonstrates exceptional learning capabilities, accurately mapping the relationship between the density profile of a carbon dioxide system to its one-body direct correlation function using an atomistic polarizable model. The neural operator model can be generalized to more complex systems, offering high-precision cDFT calculations at low computational cost.

[*]To whom correspondence should be addressed. Email: jwu@engr.ucr.edu

## I. Introduction

Classical density functional theory (cDFT) is a versatile method in statistical mechanics to describe the structure and thermodynamic properties of many-body systems based on the one-body density profiles[1-3]. Like its electronic counterpart, the practical applications of cDFT are often constrained by approximations in formulating a reliable free-energy functional and by the numerical complexity in solving the density profiles through variational minimization. Although highly accurate functionals have been developed by using the fundamental measure theory and perturbation expansions[4-7], cDFT has been predominantly applied to simple molecular systems, often represented by coarse-grained models for systems with only one-dimensional density inhomogeneity. While recent advances in machine learning methods offer new avenues for developing more accurate functionals and efficient numerical algorithms[8-17], existing studies primarily focus on simple fluids represented by the hard-sphere or Lennard-Jones models. Whereas cDFT entails operator relationships between density and potential fields, these studies mostly rely on conventional deep learning methods for functional learning[18]. In this work, we introduce a convoluted operator learning network (COLN) for mapping high-dimensional density profiles in cDFT to their corresponding one-body direct correlation functions. By applying the operator-learning method to carbon dioxide ($CO_2$) modeled as three polarizable atoms, we demonstrate that COLN has advantages over functional learning in terms of accuracy, flexibility, and scalability while also addresses the computational challenges associated with high-dimensional function mappings.

As one of the primary greenhouse gases, carbon dioxide plays a pivotal role in global warming[19]. Human activities, such as fossil fuel combustion and industrial emissions, have drastically increased atmospheric $CO_2$ concentrations, contributing to rising temperature and more frequent extreme weather events[20, 21]. Reducing $CO_2$ emissions has become central to global climate policies. In addition to carbon capture and storage (CCS) technologies for mitigating net emission, $CO_2$ holds significant industrial value from a practical perceptive. It is widely utilized in chemical production, oil and gas extraction[22-25]. $CO_2$ also serves as a raw material for producing fuels and materials, promoting the development of a carbon circular economy[26, 27]. These applications underscore the importance of fundamental research in bridging the knowledge gas between the molecular properties of $CO_2$ and its macroscopic properties. In particular, a thorough understanding of the intrinsic thermodynamic behavior of $CO_2$ is crucial for diving industrial innovation and sustainable development.

Numerous molecular models have been developed for carbon dioxide[28-35]. Among the most notable is the elementary physical model (EPM2) introduced by Harris and Yung[28]. EPM2 effectively captures the key properties of $CO_2$ by adjusting parameters in the pair potential function. Subsequent advancements have focused on incorporating bond flexibility[29], molecular polarization[30], and other many-body interactions[31], significantly improving the precision of $CO_2$ simulations. In recent years, *ab initio* molecular dynamics simulations and machine-learned potentials have also been used to predict $CO_2$ phase behavior and thermodynamic properties[36].

## II. Theoretical Models and Methods

### A. Data-Enabled Molecular Density Functional Theory

The basic concepts of classical density functional theory (cDFT) for molecular systems have been well established.[37] For a thermodynamic system consisting of $CO_2$ molecules specified by temperature $T$, chemical potential $\mu$, and volume $V$, the grand canonical potential is defined as

$$\Omega[\rho(\boldsymbol{R})] = F^{id}[\rho(\boldsymbol{R})] + F^{exc}[\rho(\boldsymbol{R})] + \int d\boldsymbol{r}\, \rho(\boldsymbol{R})[V^{ext}(\boldsymbol{R}) - \mu] \qquad (1)$$

where $\boldsymbol{R}$ represents the molecular configuration, i.e., the atomic positions of each $CO_2$ molecule, $\rho(\boldsymbol{R})$ denotes the one-body molecular density profile, $F^{id}$ and $F^{exc}$ are the ideal and excess intrinsic Helmholtz energy functionals, respectively, and $V^{ext}(\boldsymbol{R})$ is the one-body external potential. The ideal part of the intrinsic Helmholtz energy is the same as that of an ideal $CO_2$ gas with the same one-body density profile[38]

$$F^{id} = \int d\boldsymbol{R}\, \rho(\boldsymbol{R})\{k_B T \ln[\rho(\boldsymbol{R}) q_I \Lambda^3] - 1\} \qquad (2)$$

where $\Lambda = \sqrt{h^2/2\pi m k_B T}$ is the thermal wavelength, $h$ is the Planck constant, $m$ is the molecular mass, $k_B$ is the Boltzmann constant, and $q_I$ is the single-molecule partition function. In this work, $CO_2$ is modeled as a rigid linear molecule, thus the molecular configuration can be represented by the center of mass position $\boldsymbol{r}$ plus polar and azimuthal angles $(\theta, \phi)$. Accordingly, we have $\boldsymbol{R} = (\boldsymbol{r}, \theta, \phi)$ and $q_I = 4\pi$.

The excess part of the intrinsic Helmholtz energy accounts for intermolecular interactions and correlation effects. While this term is generally unknown for non-ideal molecular systems, its functional derivative with respect to the one-body density profile can be calculated from molecular dynamics or Monte Carlo simulations

$$c_1(\boldsymbol{R}) \equiv -\frac{\delta \beta F^{exc}}{\delta \rho(\boldsymbol{R})} = \ln[\rho(\boldsymbol{R}) q_I \Lambda^3] + \beta[V^{ext}(\boldsymbol{R}) - \mu] \qquad (3)$$

where $\beta = 1/k_B T$, and $c_1(\boldsymbol{R})$ is called the one-body direct correlation function. As demonstrated by Sammuller et al.[11], the intrinsic relationship between $c_1(\boldsymbol{R})$ and $\rho(\boldsymbol{R})$ can be established with machine-learning methods. Subsequently, the excess intrinsic Helmholtz energy can be obtained by functional line integration

$$\beta F^{exc} = -\int_0^1 d\lambda \int d\boldsymbol{R} \rho(\boldsymbol{R}) c_1(\boldsymbol{R}; \lambda\rho(\boldsymbol{R})). \tag{4}$$

While $F^{exc}$ is derived from various theoretical approximations in conventional cDFT methods, the data-driven approach relies on operator learning to establish the relationship between functions $c_1(\boldsymbol{R})$ and $\rho(\boldsymbol{R})$. Toward that end, we need to carry out molecular simulations to obtain high quality data for training the functional relationships. While the data-driven cDFT offers limited advantages over molecular simulation to predict the thermodynamic properties of a particular system under a specific condition, it allows for flexible applications to the same thermodynamic system under arbitrary external potentials, making it convenient for systematically exploring the thermodynamic behavior.

**B. Molecular Simulation**

The database for training the operator mapping of the one-body density profile to the one-body direct correlation function is generated through grand canonical Monte Carlo (GCMC) simulations. In this work, $CO_2$ molecules are represented by the polarizable Gaussian-charge model proposed by Jiang et al.[30] As shown schematically in Figure 1(a), the atomic model assumes a rigid linear configuration and adopts three classical Drude oscillators on the atomic sites to incorporate the polarization effects. Whereas alternative models, either *ab initio* or coarse-grained, may also be used to generate the data, it has been demonstrated that the polarizable Gaussian-charge model is highly accurate for describing both the intermolecular potential and thermodynamic properties of $CO_2$ over a wide range of conditions.[30] In Supporting Information, Figure S1 shows a comparison of various molecular models of $CO_2$ for predicting the second virial coefficient over a broad range of temperature.

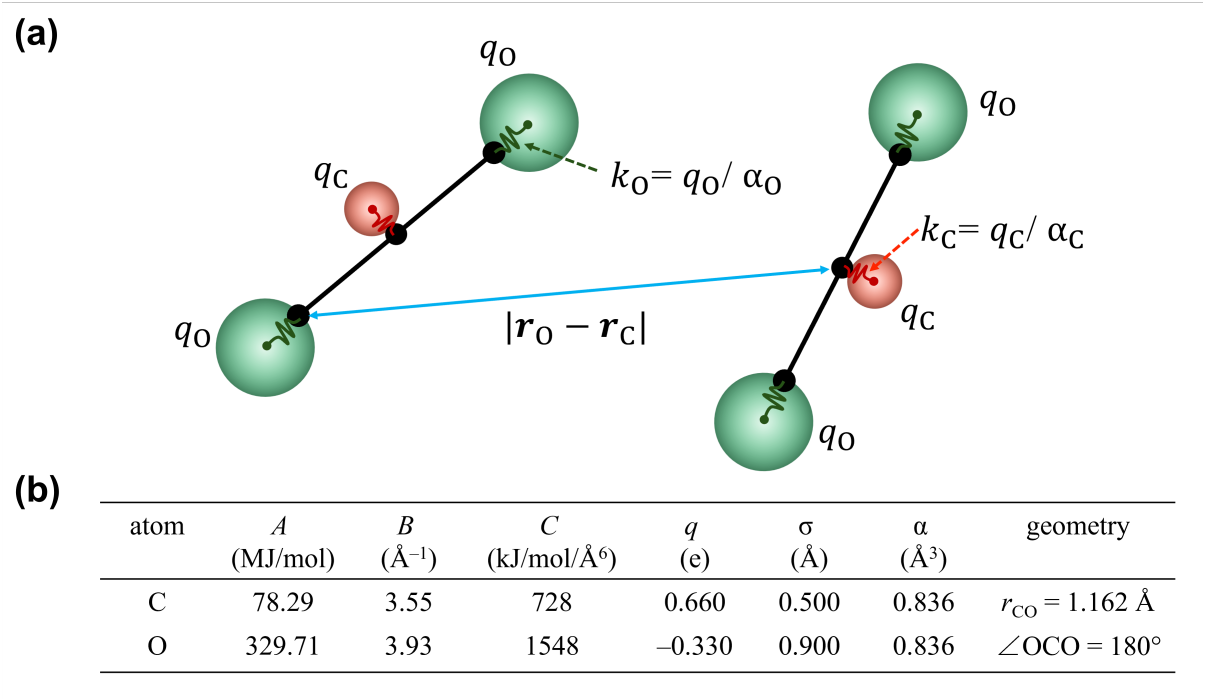

**Fig. 1** Schematic of the polarizable Gaussian-charge model for $CO_2$ (a) and the model parameters (b) describing the molecular structure, atomic polarization, and site-site interactions.

According to the polarizable Gaussian-charge model, the potential energy due to the interaction among $CO_2$ molecules can be represented by a pairwise additive interatomic potential

$$u(\mathbf{r}_i, \mathbf{r}_j) = \frac{q_i q_j}{4\pi\epsilon_0 |\mathbf{r}_i - \mathbf{r}_j|} \text{erf}\left(\frac{|\mathbf{r}_i - \mathbf{r}_j|}{\sqrt{2(\sigma_i^2 + \sigma_j^2)}}\right) + A_{ij}\exp(-B_{ij}|\mathbf{r}_i - \mathbf{r}_j|) - C_{ij}/|\mathbf{r}_i - \mathbf{r}_j|^6 \quad (5)$$

where $q_i$ and $q_j$ are the Gaussian charges of atoms $i$ and $j$[39], $|\mathbf{r}_i - \mathbf{r}_j|$ denotes the distance between the two atoms, $\sigma_i$ and $\sigma_j$ are the widths of the charge distributions of the respective atoms. In Eq. (5), parameters $A_{ij}$, $B_{ij}$, and $C_{ij}$ are atom-specific and have the same meanings as those appeared in the Buckingham exp-6 potential. These parameters, along with those related to the atomic charge, polarization, and molecular geometry, are presented in Figure 1(b). The cross-interaction parameters $A_{ij}$ and $B_{ij}$ are predicted from the Kong-

Chakrabarty combining rules

$$A_{ij} = \frac{1}{2}\left[A_{ii}\left(\frac{A_{ii}B_{ii}}{A_{jj}B_{jj}}\right)^{-B_{ii}/(B_{ii}+B_{jj})} + A_{jj}\left(\frac{A_{jj}B_{jj}}{A_{ii}B_{ii}}\right)^{-B_{jj}/(B_{ii}+B_{jj})}\right] \quad (5a)$$

$$B_{ij} = \frac{2B_{ii}B_{jj}}{B_{ii}+B_{jj}} \quad (5b)$$

and $C_{ij}$ is obtained from the geometric mixing rule, $C_{ij} = \sqrt{C_{ii}\,C_{jj}}$.

The training data for $c_1(\boldsymbol{R})$ and $\rho(\boldsymbol{R})$ were generated by carrying out GCMC simulation for $CO_2$ under supercritical conditions (Figure 2). For each atomic site, the external potential varies only in a single direction, which is designated as $x$. Accordingly, $\rho(\boldsymbol{R})$ and $c_1(\boldsymbol{R})$ can be expressed in terms of 3-dimensional functions, $\rho(x,\theta,\phi)$ and $c_1(x,\theta,\phi)$, respectively. The atomic external potential consists of multiple sets of sinusoidal functions with different frequencies and phases, plus an equal number of discontinuous linear functions varying in a single direction of the simulation space[11]

$$V^{ext}(x)/k_B = \sum_{n=1}^{4} A_n T \sin\left(\frac{2\pi n x}{L} + \phi_n\right) + V_n(x) \quad (6)$$

where $A_n$ and $\phi_n$ are Fourier coefficients and phase angles, $L$ is the length of the simulation box in the $x$ direction, and the discontinuous linear functions have the form

$$V_n(x) = \begin{cases} V_{n1} + \dfrac{x - x_{n1}}{x_{n2} - x_{n1}}(V_{n2} - V_{n1}), & if \quad x_{n1} < x < x_{n2} \\ 0, & otherwise \end{cases} \quad (7)$$

where $x_{n1}$ and $x_{n2}$ are uniformly chosen between 0 and $L$, $V_{n1}$ and $V_{n2}$ are drawn from a normal distribution with a mean value of $-T$ and a variance of $T$, the latter represents the absolute temperature. In Eq. (6), $A_{n=1,2,3,4}$ are randomly selected from a normal distribution with a zero mean and a variance of 1, and $\phi_{n=1,2,3,4}$ are from a uniform distribution over the interval [0, 2π).

The simulation box has the dimensions of 40 Å × 40 Å × 40 Å, with the periodic boundary conditions applied in all spatial directions. The GCMC calculations were conducted over eighty thermodynamic conditions for the bulk gas. To avoid potential phase transitions, we take supercritical temperatures, $T = 400, 500, 600, 700, 800, 900, 1000$ and $1200$ K, and a range of reduced chemical potentials, $\beta\mu = -\{14, 15, 16, 17, 18, 19, 20, 21, 22, 23\}$. In addition, we set $V^{ext} = \infty$ for $|L/2 - x| > |L/2 - x_w|$, with $x_w$ randomly selected from the interval of 2 to 6 Å, to effectively create a hard wall at the edge of the simulation domain. Additional details of the GCMC simulation are described in Supporting Information (SI).

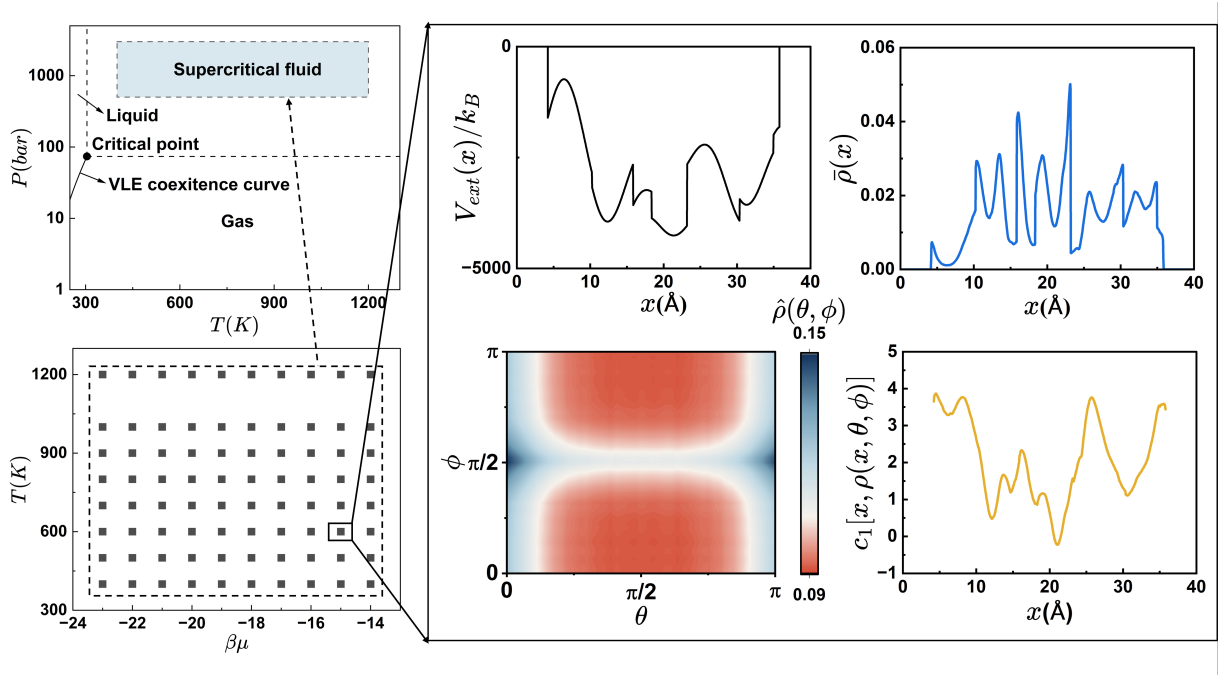

**Fig. 2 The training data obtained from GCMC simulation.** The diagram on the left illustrates the simulation parameter space and the corresponding phase diagram. The plots on the right highlight one specific set of data obtained from GCMC simulation.

In data-driven cDFT, the primary purpose of molecular simulation is to obtain the training data for the molecular density profiles, $\rho(x, \theta, \phi)$, under different thermodynamic conditions and external potentials. To reduce the dimensionality of the molecular density profile, we have also sampled the angle-averaged local density $\bar{\rho}(r)$, and the position-averaged angular function $\hat{\rho}(\theta, \phi)$, as shown in Figure 2. It is important to note that for $\hat{\rho}(\theta, \phi)$, we only need to calculate its value in the $\phi$ direction within the range of [0, π]. This is because the density profile exhibits a periodicity of π in the $\phi$ direction, which is smaller than the intrinsic periodicity of angular coordinate systems.

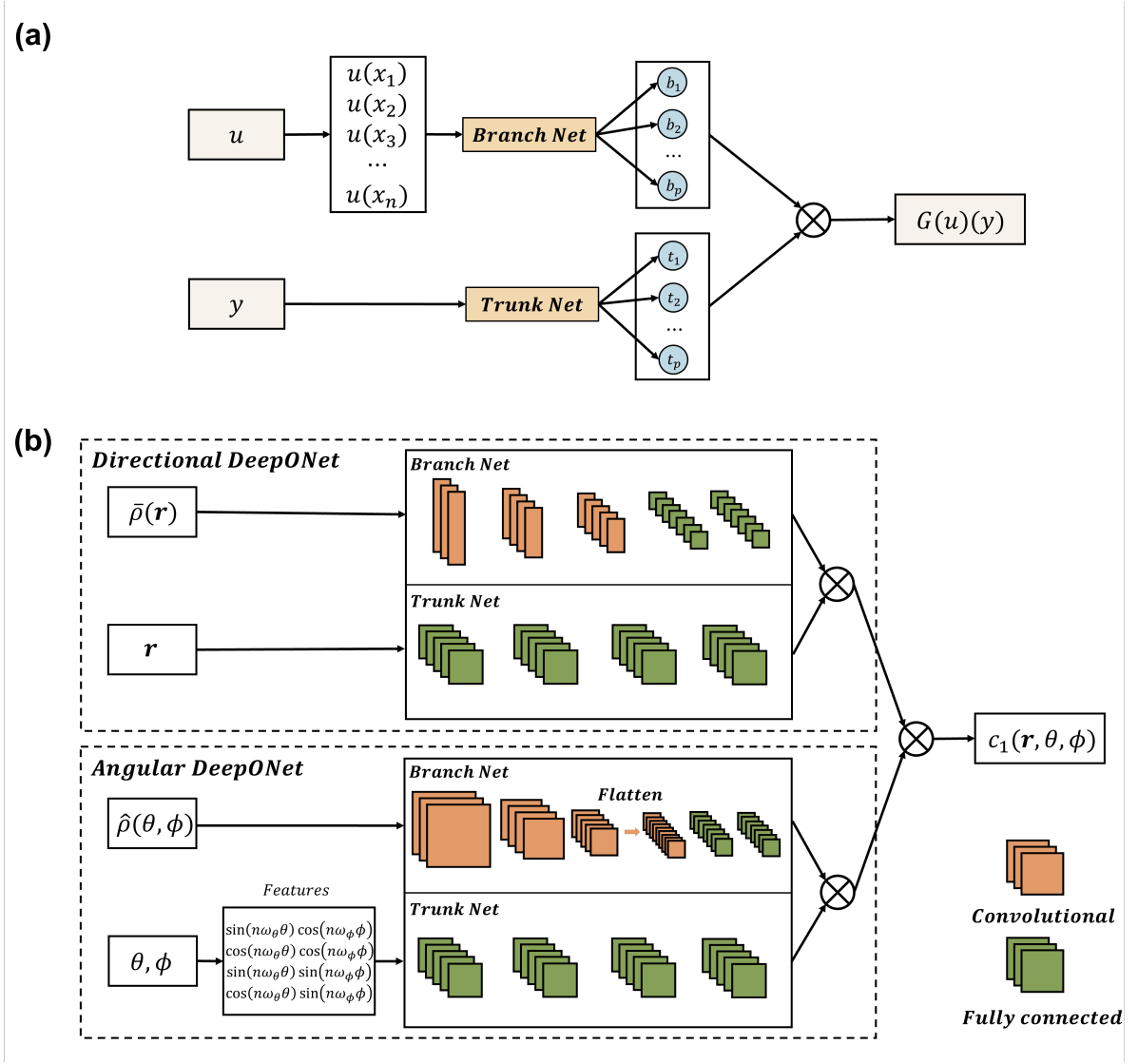

**Fig. 3 Illustration of the deep operator learning networks.** (a) The general architecture of the "vanilla" DeepONet; (b) The convoluted operator learning networks (COLN) proposed in this work.

## C. Convoluted Operator Learning Network (COLN)

Our operator learning method is inspired by the deep operator network (DeepONet) model to approximate nonlinear continuous operators.[40] The DeepONet architecture is grounded in a rigorous mathematical framework known as the universal operator approximation,[41] which asserts that a neural network with a single hidden layer can accurately approximate any nonlinear continuous operator.

Figure 3(a) presents the general architecture of the "vanilla" DeepONet, a basic form without any additional modifications or enhancements. The machine-learning model consists of two sub-neural networks, referred to as the branch net and the trunk net. For learning an operator with input function $u(x)$ and output function $G(y, u(x))$, the branch net samples $n$ scattered points at fixed locations for each input, denoted as $u(x_1), u(x_2), \ldots, u(x_n)$. Meanwhile, the trunk net takes a number of scattered points for $y$, the independent variable of the output function. The final output of the operator learning is obtained by a dot product of the outputs of these two sub-networks

$$G(y, u(x)) = \sum_{k=1}^{p} \underbrace{b_k(u(x_1), u(x_2), \ldots, u(x_n))}_{branch} \underbrace{t_k(y)}_{trunk} \qquad (8)$$

where $p$ represents the dimension of the output vectors from the branch and trunk nets.

Intuitively, the branch-trunk structure can be interpreted as a trunk network, where each weight in the final layer is parameterized by a separate branch network. When the DeepONet is applied to multidimensional mappings, the input to the branch net must also be multidimensional. Even when the dimensionality of each input is relatively low, the combined input $u(x_a, x_b, x_c)$ can become computationally infeasible, where subscripts $a, b$ and $c$ denote the dimensions of each input. For a moderate 3D problem with $a, b, c = 64$, even the latest NVIDIA Ampere GPUs lack the sufficient memory to process a single training sample.[42] In Supporting Information, we will demonstrate the specific structure of vanilla DeepONet when applied to solve multidimensional problems in classical DFT. Because of the computational challenges, we do not directly use the vanilla DeepONet to establish operator mapping from $\rho(\boldsymbol{R})$ to $c_1(\boldsymbol{R})$.

For systems consisting of linear molecules such as $CO_2$, both the molecular density profile and the one-body direct correlation function depend on molecular position and orientation. In this case, we may expand $\rho(\boldsymbol{r},\theta,\phi)$ and $c_1(\boldsymbol{r},\theta,\phi)$ in terms of the spherical harmonics $Y_{ml}(\theta,\phi)$:

$$\rho(\boldsymbol{r},\theta,\phi) = \sum_{l=0}^{\infty}\sum_{m=-l}^{l} g_{ml}(\boldsymbol{r})Y_{ml}(\theta,\phi) \tag{9}$$

$$c_1(\boldsymbol{r},\theta,\phi) = \sum_{l=0}^{\infty}\sum_{m=-l}^{l} c_{ml}(\boldsymbol{r})Y_{ml}(\theta,\phi) \tag{10}$$

where $g_{ml}(\boldsymbol{r})$ and $c_{ml}(\boldsymbol{r})$ are coefficients (viz., "projections") of the expansion. Instead of directly mapping the relationship between $\rho(\boldsymbol{r},\theta,\phi)$ and $c_1(\boldsymbol{r},\theta,\phi)$, $c_1(\boldsymbol{r},\theta,\phi)$ could be alternatively obtained from its spherical harmonic coefficients

$$c_1\big(\boldsymbol{r},\theta,\phi,\rho(\boldsymbol{r},\theta,\phi)\big) = \sum_{l=0}^{\infty}\sum_{m=-l}^{l} c_{ml}\big(\boldsymbol{r},\rho(\boldsymbol{r},\theta,\phi)\big)Y_{ml}(\theta,\phi) \tag{11}$$

Although $c_{ml}\big(\boldsymbol{r},\rho(\boldsymbol{r},\theta,\phi)\big)$ entails a multi-dimensional function $\rho(\boldsymbol{r},\theta,\phi)$ as the input, Eqs. (9-11) suggest that the position and angular variables can be deconvoluted by using two DeepONets of lower dimensionality. As shown in Figure 3(b), one DeepONet uses the locally angle-averaged density profile $\bar{\rho}(\boldsymbol{r})$ as the input, and the other depends on the position-averaged angular function $\hat{\rho}(\theta,\phi)$. A major advantage of this convoluted operator-learning network is that it allows us to isolate the periodic angular components from the position-dependent quantities. While $\bar{\rho}(\boldsymbol{r})$ generally does not exhibit periodic characteristics, we can apply the periodic boundary conditions specifically to $\hat{\rho}(\theta,\phi)$, thereby enhancing the training performance. In Supporting Information, we provide additional details for the architecture and implementations of the directional DeepONet and the angular DeepONet.

**III. Results**

We implemented our convoluted operator learning network using the open-source PyTorch framework[43]. After testing various common activation functions, we found that LeakyReLU is particularly effective in mitigating the gradient vanishing problem while also facilitating a further reduction in the loss function. Furthermore, we employed Kaiming normal initialization[44], which is optimized for ReLU-based activation functions, to prevent the machine-learning model from converging to local minima and to enhance the overall convergence efficiency. The standard Adam optimizer[45] is used for training the model parameters. To mitigate the risk of overfitting, we incorporate a dropout mechanism[46] into the machine-learning model, setting the dropout rate at 0.5. Furthermore, the Adam optimizer is configured with a weight decay of $1 \times 10^{-4}$, serving as a regularization technique to prevent overfitting by constraining the network weights.

One data point is represented as a quintuplet, denoted as $(\bar{\rho}(x), x, \hat{\rho}(\theta, \phi), (\theta, \phi), c_1(x, \theta, \phi))$, where $\bar{\rho}(x)$ is obtained by averaging $\rho(x, \theta, \phi)$ over all orientations at position $x$, which is sampled within a range of 24 Å starting from a random point in one set of the density profiles. The exact value of $x$ is randomly selected from the directional sampling range. Meanwhile, $\hat{\rho}(\theta, \phi)$ is obtained by using the same averaging method but within the angular slice under consideration and sampled on an equal-spaced grid of $(\pi, \pi)$ with a $\pi/30$ gap in each dimension. Additionally, the value of $(\theta, \phi)$ is randomly chosen from the angular sampling range, and $c_1(x, \theta, \phi)$ is randomly sampled from a combined space of the directional and angular dimensions, but with a shorter sampling range of 12 Å in the $x$ direction. The reduction of the sampling range is necessary because the mapping we intend to learn, $\rho(x,\theta,\phi) \to c_1(x,\theta,\phi)$, is inherently global.

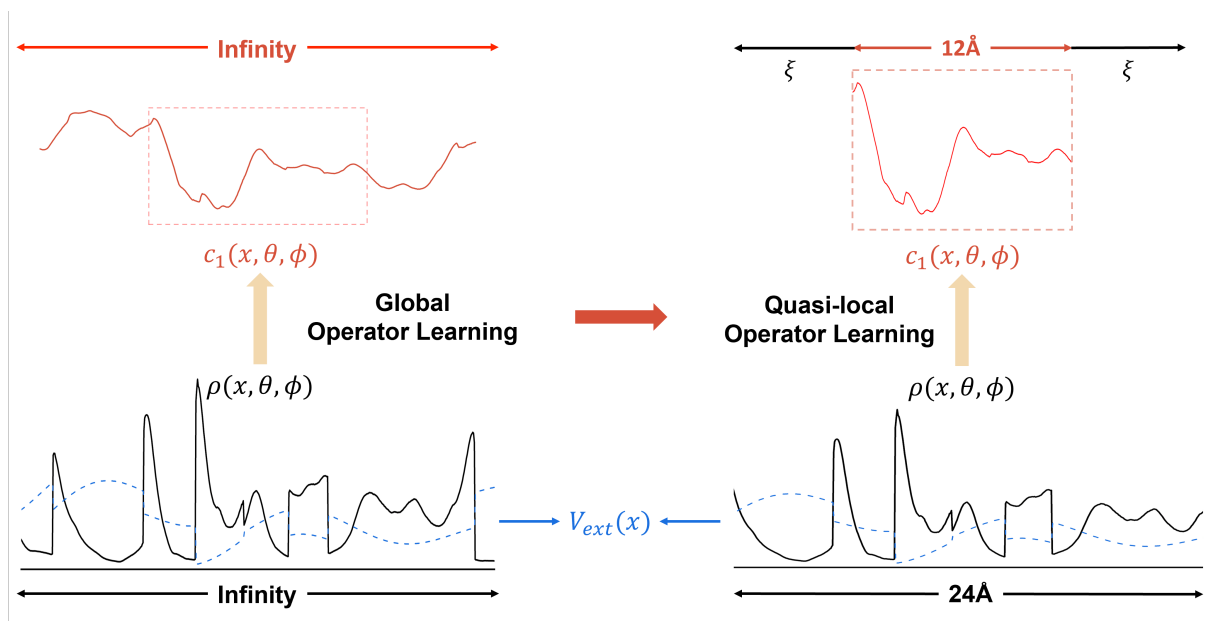

**Fig. 4 Illustration of the global and quasi-local operator learning schemes.** The diagram on the left shows the global operator learning, and the diagram on the right shows the quasi-local operator learning scheme.

As shown in Figure 4, we used a quasi-local operator learning scheme for training the functional relationship between $\rho(x,\theta,\phi)$ and $c_1(x,\theta,\phi)$. A similar strategy was used by Sammuller et al.[11] but for training $c_1(x,\theta,\phi)$ as a neural functional at each position. Here, we selected a quasi-local system of 12 Å length in the $x$ direction and chose the local density information within the correlation length $\xi$ beyond the discussed local range as supplementary input to mitigate the discrepancies between local and global mappings. The quasi-local mapping is justified because the correlation length represents the effective distance of molecules in the system interacting with each other. Specifically, the correlation length employed in quasi-local method is set to 6 Å, which is approximately 2 to 3 times the size of a carbon dioxide molecule, and further investigation indicates that extending the correlation length does not lead to any significant improvement in the performance of the trained model. It should be noted that $c_1(x,\theta,\phi)$ cannot be directly determined from molecular simulation when the local density vanishes, $\rho(x,\theta,\phi) = 0$. As shown in Eq. (3), this occurs within the hard-wall regions, excluding the direct sampling of $c_1(x,\theta,\phi)$ at positions where the external potential diverges. However, the zero-density regions are necessary for $\rho(x,\theta,\phi)$ sampling, as they enable the trained model to recognize the presence of hard walls established in the simulation. The inclusion of zero-density data enhances the accuracy of the predicted density profile, as discussed in the following sections.

Considering the inherent mirror symmetry of the operator mapping between $\rho(x,\theta,\phi)$ and $c_1(x,\theta,\phi)$, we apply a mirror flip operation to the 320 data sets generated by the GCMC simulation, effectively doubling the size of our database. Recognizing that one data point is a quintuplet as described above, we can select multiple different values of $x$ and $(\theta,\phi)$ for each density profile thereby also effectively expanding our database. Additionally, we applied Min-Max normalization to the database using the normalized variable

$$u_{nor} = \frac{u - u_{Min}}{u_{Max} - u_{Min}} \tag{12}$$

The procedure not only enhances the credibility of the test error but also benefits the overall training process. Furthermore, it ensures that the input values are scaled uniformly, thereby facilitating more efficient convergence during model training. In total, $6.4 \times 10^5$ pairs of data points are generated from the 320 sets of original data. This extended dataset significantly contributes to the robustness of our model by providing a comprehensive range of training examples.

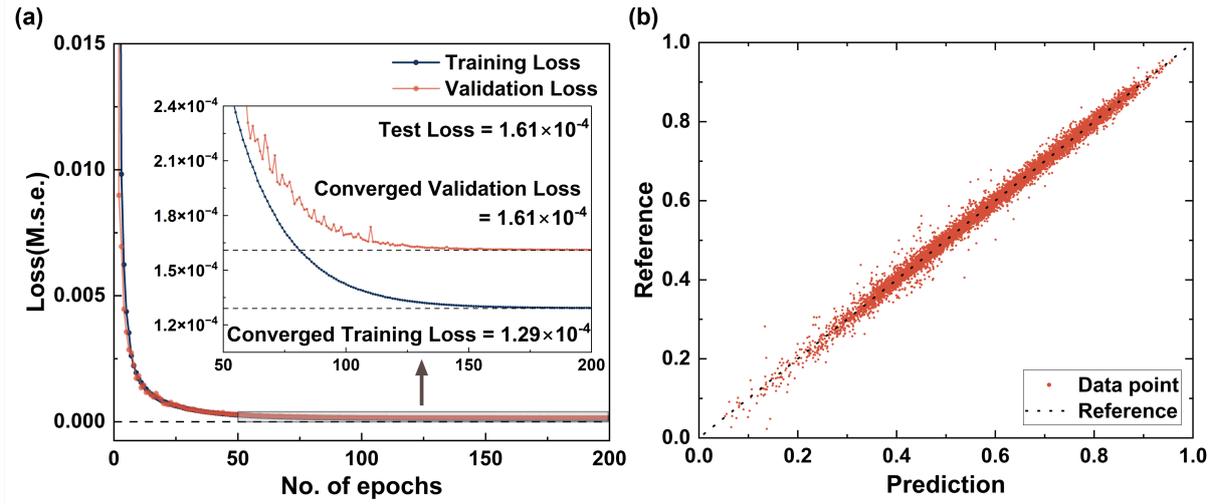

**Fig. 5 Illustration of the training result.** (a) The evolution of training and validation loss over the number of epochs; (b) Comparison of the reference and model prediction.

Prior to training our machine-learning model, we followed a conventional practice to partition the total database into three subsets: the training set, validation set, and test set, comprising 80%, 10%, and 10% of the total database, respectively. The training set is utilized for parameter optimization of the machine-learning model, while the validation set is employed to compute validation errors after each epoch, enabling us to monitor the training process and assist in hyperparameter tuning. The test set, on the other hand, is excluded from the training process and is exclusively reserved for evaluating test errors after the completion of training, where the loss function used above is the mean squared error (m.s.e.) between the true value of $c_1(x, \theta, \phi)$ and the model prediction. We train the model for 200 epochs, starting with an initial learning rate of 0.001, which is subsequently reduced to 0.96 times its original value after each epoch. This approach aids the training process in achieving smoother convergence, as illustrated in Figure 5(a). Given an initial learning rate of 0.001, as is convention, we conduct extensive tests by varying the learning rate decay within the interval [0.9, 0.99]. We use the number of epochs required to reach a convergence accuracy of $1 \times 10^{-4}$ as the primary optimization criterion and the training parameters described above have been proven to be the optimal setup, achieving smooth convergence without compromising accuracy.

Figure 5(a) shows the evolution of training and validation loss over the number of epochs. The consistency between the training and validation losses indicates that overfitting has not occurred, as both losses converge closely. The training loss stabilizes at $1.29 \times 10^{-4}$, while the validation loss converges slightly higher at $1.61 \times 10^{-4}$, matching the test error. This suggests that the machine-learning model generalizes well to unseen data. The operator learning model prediction from test set is illustrated in Figure 5(b), which shows high consistency between the prediction and reference labels.

To test the performance of the operator learning for the thermodynamic model, we first utilize the trained neural operator model to predict the direct correlation function over the entire domain. Figure 6(a) and (b) presents the prediction of $c_1(x, \theta, \phi)$ with either the angular variables or the directional variable fixed, respectively. In Figure 6(a), 6 sets of $(\theta_0, \phi_0)$ are randomly selected in the range of $[0, \pi]$, i.e., $[(\theta_0, \phi_0) = (19\pi/30, \pi/6), (2\pi/3, \pi/5), (9\pi/10, \pi/6), (11\pi/15, \pi), (23\pi/30, 11\pi/30), (11\pi/30, 3\pi/10)]$. In Figure 6(b), $x_0 = 8.48$Å is randomly selected from $x \in [0, 24 \text{ Å}]$. While the position-averaged angular function exhibited notable discrepancy at certain ranges of molecular angles, the machine-learning model shows excellent overall predictive capabilities. In particular, the test loss consistently stayed within a narrow range of $1 \times 10^{-4}$, indicating the robust performance. The excellent performance of our machine learning model in predicting the one-body direct correlation function over the multidimensional continuous space indicates that it is not merely a local interpolation scheme based on the training data, but it has truly learned the mapping relationship between $\rho(x, \theta, \phi)$ and $c_1(x,\theta,\phi)$. This suggests that, a surrogate model of carbon dioxide is established under the conditions specified by the provided simulation data and, in principle, it is capable to predict the intrinsic thermodynamic properties of carbon dioxide under arbitrary conditions within the theoretical framework of cDFT.

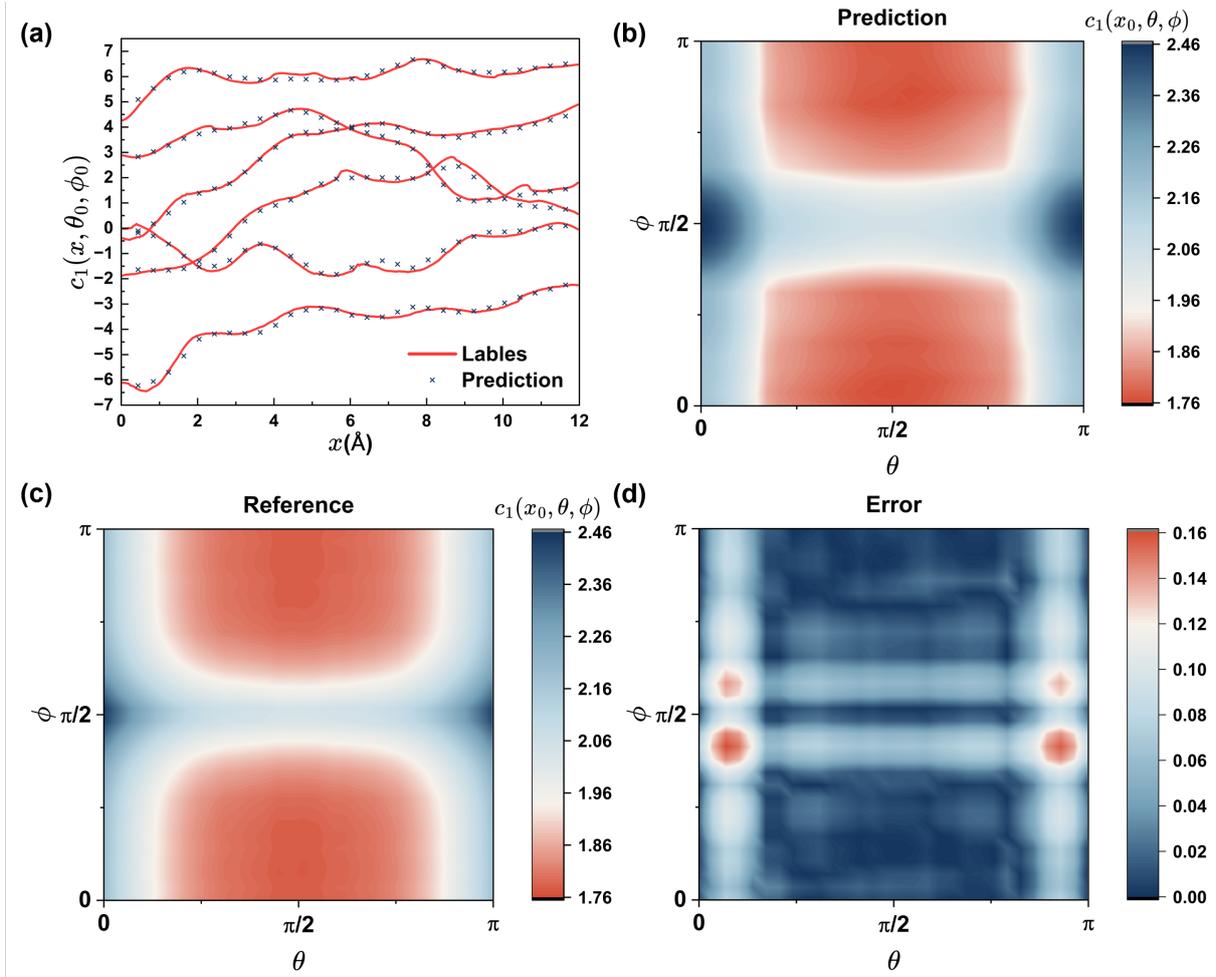

**Fig. 6 The predictions of one-body direct correlation functions** (a) The position component of $c_1(x,\theta,\phi)$ at 6 sets of fixed angular variables $(\theta_0,\phi_0)$ randomly selected in the range of $[0,\pi]$; (b) The angular term $(\theta,\phi)$ at fixed directional variable $x_0 = 8.48\text{Å}$; (c) The reference of (b); (d) Absolute error of the prediction on angular plane.

To further validate our neural-operator model, we predicted $\rho(x,\theta,\phi)$ using the Euler–Lagrange equation

$$\rho(x,\theta,\phi) = \frac{1}{4\pi\Lambda^3}\exp[c_1(x,\theta,\phi,\bar{\rho}(x),\hat{\rho}(\theta,\phi)) + \beta(\mu - V_{ext}(x))] \qquad (13)$$

Given an expression for the reduced one-body potential $\beta(\mu - V_{ext}(x))$ along with the operator relationship for $c_1(x,\theta,\phi)$, we can determine the density profiles from Eq. (13) self-consistently by using the Picard iteration. Considering the long-range nature of density correlations induced by intermolecular interactions, it is crucial to implement the quasi-local approximation with sufficient width in predicting the molecular density profile. To account for the boundary effects, we choose $\xi$ as the width of $c_1(x,\theta,\phi)$ within the hard wall where the molecular density profile vanishes. The extended density profile effectively cuts off the influence of the unseen potential beyond the domain under consideration. The specific scheme is to set $V_{ext}(x) = \infty$ and $\rho(x,\theta,\phi) = 0$ for $|x - x_w| < \xi$, where $x_w$ denotes the distance to the hard-wall boundary.

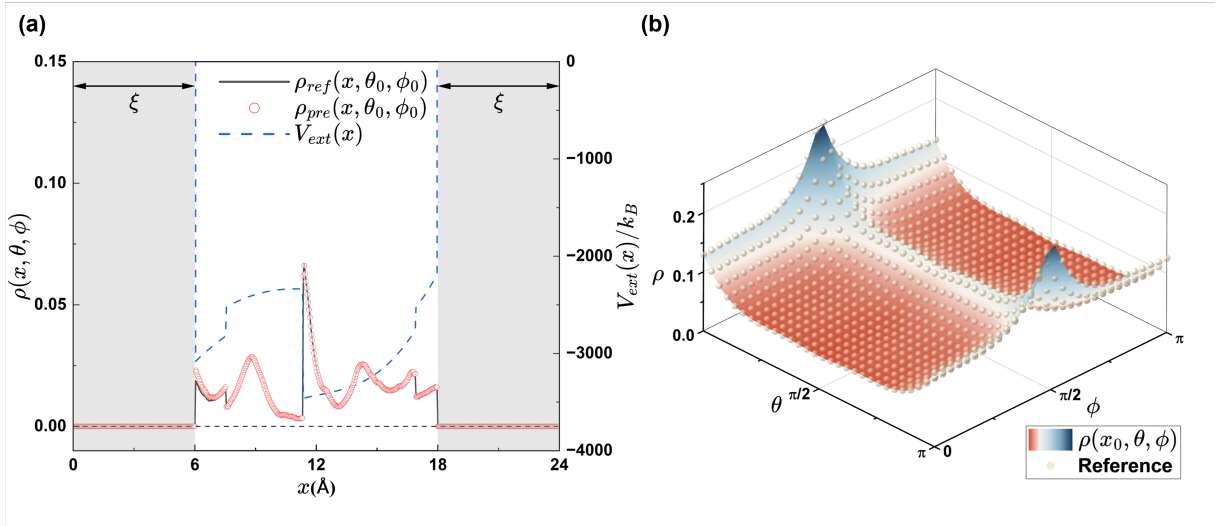

**Fig. 7 The predictions of density profile from the data-enabled molecular cDFT.** (a) The predicted density $\rho(x,\theta,\phi)$ with angular variables fixed and (b) the predicted density $\rho(x,\theta,\phi)$ with directional variables fixed. The temperature and chemical potential of the system are randomly chosen as $400K$ and $-16/\beta$.

As shown in Figure 7, the neural operator model yields accurate prediction of $\rho(x,\theta,\phi)$

under an arbitrarily selected external potential. Here, the angular variables are randomly fixed to $(\pi/5, 2\pi/3)$, and the directional variable is randomly fixed to 12 Å. In both cases, the density profiles predicted by cDFT over the entire spatial domains are highly accurate, with the overall average errors converged around $5.6 \times 10^{-5}$ after $L_2$-normalization. The conventional molecular simulation scales $O(n)$ for systems with truncated pairwise potentials, and $O(nlog(n))$ for simulating systems with long-range interactions, where $n$ is the number of atoms or molecules.[47] However, the machine-learning technique scales $O(1)$, implying that the computational cost is the same for either large or small systems, depending only on the length scale. Therefore, compared to GCMC simulations, the neural-operator model can predict the equilibrium density profile with negligible computational cost. Importantly, the machine-learning model is applicable to the carbon dioxide system under any external potential. Thus, we demonstrate that the neural operator learning can replace conventional approaches to formulate the free-energy functional that integrates the high accuracy of molecular simulation with the high efficiency of cDFT calculations. The computational efficiency makes the data-enabled cDFT a practical and scalable approach without compromising on precision. Given the broad applicability of dimension reduction techniques to diverse molecular systems, we expect that the operator learning scheme will unlock new avenues for a broader application of cDFT.

**IV. Conclusion**

In this work, we proposed a convoluted operator learning network (COLN) for mapping the relationship between the molecular density profile and the one-body direct correlation function of carbon dioxide using a polarizable atomic model. The machine-learning model leverages GCMC simulations and cDFT calculations to provide accurate and efficient predictions for complex molecular systems, enabling the exploration of diverse thermodynamic regimes. Future work could concentrate on utilizing of the operator learning method to predict thermodynamic properties associated with $CO_2$ capture and conversion, thereby broadening the applicability beyond traditional simulation techniques. Additionally, enhancements in the accuracy of the databases used for operator learning could be explored, with first-principles-based molecular simulations emerging as a promising avenue for further improvement.